\documentstyle[12pt]{article}
\def\overlay#1#2{\setbox0=\hbox{$#1$}\setbox1=\hbox to \wd0{\hss$#2$\hss}#1%
\hskip -1\wd0\copy1}

\def\bra{\langle}
\def\ket{\rangle}

\def\eea{\end{eqnarray}}
\def\bea{\begin{eqnarray}}
\def\eeas{\end{eqnarray*}}
\def\beas{\begin{eqnarray*}}
\def\ee{\end{equation}}
\def\be{\begin{equation}}

\def\bold#1{\setbox0=\hbox{$#1$}%
      \kern-.025em\copy0\kern-\wd0
      \kern.05em\copy0\kern-\wd0
      \kern-.025em\raise.0433em\box0 }

\begin{document}

%
\hfill hep-ph/9310250

\title{ \bf STRONG DECAYS OF $Q \bar Q$ MESONS}
%
\renewcommand{\thefootnote}{\fnsymbol{footnote}}

\author{\large Carlo Gobbi$^*$,
         Francesco Iachello and Dimitri Kusnezov\\   \\
         Center for Theoretical Physics, Sloane Physics Laboratory, \\
         Yale University, New Haven, Connecticut 06511 \\ }
\date{}

\maketitle

\vspace{25pt}
\begin{abstract}
We present a detailed study of the two--body strong decays of light mesons.
Both the space part and the spin--flavor--color part of the wave functions
are generated algebraically and closed forms are obtained for all decays.
Experimental deviations from our systematics are seen to be suggestive
of both missing mesons and exotic QCD configurations.
\end{abstract}

\vspace{2cm}


\newpage

\section{Introduction}
\hspace*{0.6cm}
In the last 30 years, a considerable amount of experimental information
has been accumulated on the spectroscopy of hadrons. In addition to masses,
electromagnetic, weak and strong decay widths have been measured to the extent
that the most recent compilation \cite{PDB} contains many
pages and a voluminous
listing of decay modes of mesons and baryons. While the earlier data
were analyzed in terms of symmetries, most notably the ``eightfold" $SU(3)$
flavor symmetry of Gell--Mann~\cite{GM} and Ne'eman~\cite{Nee}, in recent
years there has been a tendency to analyze those data in terms of the
non--relativistic~\cite{QM} or relativized~\cite{GI} quark model, in which a
Schr\"odinger--like equation is solved with some two--body quark--quark
interaction $V_{ij}$. \par
We have recently initiated a reanalysis of hadronic spectroscopy by
reintroducing the concept of symmetry, now enlarged from flavor
$SU_f(3)$~\cite{GM,Nee} or flavor--spin $SU_{fs}(6)\supset SU_f(3) \otimes
SU_s(2)$~\cite{GR} to include the space part of the hadronic wave
function~\cite{IM,ID}. We use an algebra $\cal G$ that generates all
excitations, ${\cal G}={\cal R}\otimes {\cal G}_s \otimes {\cal G}_f
\otimes {\cal G}_c$, where ${\cal R}$ is the space algebra, ${\cal G}_s $
the spin algebra, ${\cal G}_f $ the flavor algebra and ${\cal G}_c$ the
color algebra. The spin--flavor--color part is the same as in earlier analyses.
The novel part is the use of a spectrum generating algebra for the space part
which allows us to construct all states starting from the ground state.
Hence we are able to put into a single representation of the algebra ${\cal G}$
all hadron states, including orbital and radial excitations and not just states
with a given value of the orbital and radial quantum number. \par
The purpose of our reanalysis is twofold:
\begin{description}
\item[(i)] we want to condense the information contained in the
extensive tabulations of Ref.\cite{PDB} into a small set of parameters;
\item[(ii)] more importantly, we want to identify ``new" physics, if any, which
emerges from the data set. If a data point deviates considerably from the
parameterization of (i) (i.e. the symmetry is badly broken), and this
deviation cannot be explained in a reasonable way, we take it as indication
that new physics is at play.
\end{description}
By new physics, in the present context,
we mean unconventional configuration of quarks and gluons. These
unconventional configurations are in many cases demanded by QCD and it would
be quite surprising if they are not found (or a reasonable explanation
is not found for their absence). An example of these configurations are,
in meson spectroscopy, purely gluonic states (glueballs). Unless a somewhat
model independent framework is set up, against which the data can be compared,
it would not be possible to identify uniquely those states which occur in a
region of overlapping resonances. \par
In this article, we analyze strong two--body decays of $q \bar q$ mesons,
and show that the known decay widths can be reasonably well summarized
by a transition operator with
only two parameters.
This parameterization provides a description of the known
widths within a factor of two or better, and thus we believe that it can make
predictions of unknown widths with this accuracy.
It would be of great interest to check these predictions by measuring
some of the unknown widths. We find it particularly important that, by
using space symmetries, the results for the decay widths can be written
in a transparent way which isolates the various contributions: kinematic
factors, spin--flavor part, and space part (form factors). The
space symmetry can be viewed as a way to generate the hadronic form factors
in a consistent way.

\section{Method of Calculation}
\hspace*{0.6cm}
In this article, we consider all two--body decays of $q \bar q$ mesons
where one of the emitted mesons is a pseudoscalar:
\be \label{dec}
M \longrightarrow M' + M'' .
\ee
In order to compute the strong decays (\ref{dec}), we need the meson wave
functions and the form of the transition operator.

\subsection*{A. Meson wave function}
We use an algebraic construction of the wave functions.
The spectrum generating algebra (SGA) is
\be
{\cal G}={\cal R}\otimes {\cal G}_s \otimes {\cal G}_f \otimes {\cal G}_c \, ,
\ee
where $\cal R$ is the space part, ${\cal G}_s$ the spin part,
${\cal G}_f$ the flavor part and ${\cal G}_c$ the color part. Following
Refs.~\cite{IM,ID} we take
\be
{\cal R}   \equiv U(4), \quad {\cal G}_s \equiv SU_s(2), \quad
{\cal G}_f \equiv SU_f(3), \quad
{\cal G}_c \equiv SU_c(3), \quad
\ee
and consider only light quarks, $u$, $d$, $s$. The corresponding wave functions
are of the type
\be
\psi=\psi_{\cal R} \otimes \psi_s \otimes
\psi_f \otimes \psi_c  \, .
\ee
The color part of $\psi$ does not play any role for $q \bar q$ mesons,
as long as we construct color singlet states. The flavor part can be written
either explicitly as $\vert q_i \bar q_j \ket$ for quark of flavor $i$
and antiquark of flavor $j$, or as a $SU_f(3)$ wave function
\be
\left \vert \matrix{SU_f(3) & \supset & SU_I(2) & \otimes & U_Y(1) &
                               \supset & SO_I(2) & \otimes & U_Y(1) \cr
                     \downarrow &      & \downarrow &      & \downarrow &
                                       & \downarrow &      &        \cr
                   (\lambda, \mu) & &   I        &         &  Y     &
                                       &    I_3     &      &          }
\right \rangle \, .
\ee
The space and spin parts are coupled to total angular momentum $J$, and
written as
\be
\vert N,f,L,S,J,M_J \ket = \sum_{M_S,M_L}
\bra L,M_L,S,M_S \vert J,M_J \ket \, \vert S,M_S\ket \, \vert f,L,M_L\ket \, .
\ee
Here $f$ is a radial quantum number, $N$ labels the irreducible
representation of $U(4)$ (i.e. the Hilbert space on which calculations are
performed) and $L$, $S$, $J$, $M_J$ have obvious meaning.
\par
In order to study the decay widths in as much as possible model independent
way,
we consider two limiting cases of wave functions corresponding to the two
dynamic symmetries of $U(4)$
\be
\left. \matrix{       &     & U(3) &          &  & &   & \quad (I)\, ,\cr
                      & \nearrow & & \searrow &  & &   & \cr
                U(4)  &     &      &    & SO(3) & \supset & SO(2) & \cr
                      & \searrow & & \nearrow &  & &   &  \cr
                      &     & SO(4) &         &  & &   & \quad (II)\, .}
\right.
\ee
The first corresponds to the harmonic--oscillator quark model~\cite{QM},
the second to a string--like situation~\cite{IM,ID}. The wave functions
of chain $(I)$ are labelled by
\be \label{i}
\vert N, n,L,M_L \ket  \, ,
\ee
where $n$ is the harmonic oscillator principal quantum number. The wave
functions of chain $(II)$ are labelled by
\be \label{ii}
\vert N, v, L, M_L \ket \, ,
\ee
where $v$ is the vibrational quantum number. The wave functions (\ref{i})
and (\ref{ii}) differ in their radial part but have the same angular part,
since
 $SO(3)$ is a common subalgebra of both $U(3)$ and $SO(4)$. The quantum number
assignments of some mesons are shown in Table I.
Another possible notation is the spectroscopic notation $^{2S+1}\! L_J$.

\subsection*{B. Form of the transition operator}

We assume that the meson $M''$ is emitted by quark $q_i$ or antiquark $\bar
q_j$
in meson $M$ which then changes to meson $M'$.
In this picture, the quark contribution to the transition operator is
of the form~\cite{GI}
\be  \label{op}
{\cal H}'=X^{M''}_{q_i q'_{i}} \, \Bigl[ \,
g \, (\bold \sigma_i \cdot \bold k)\,
e^{- i \, \bold k \cdot \bold r_i} \, + \, h \,
(\bold \sigma_i \cdot \bold p_i ) \, e^{- i \, \bold k
 \cdot \bold r_i} \, \Bigr] \, ,
\ee
where $\bold k$ is the momentum of the emitted meson, and
$X^{M''}_{q_i q'_i} $ an $SU(3)_f$ flavor matrix corresponding to quark $q_i$
changing to quark $q'_i$ with emission of meson $M''$ (Fig.1).
The second term in Eq.(\ref{op}) is written in an unsymmetrized form
since by commuting $\bold \sigma \cdot \bold p$ with  $e^{- i \, \bold k
 \cdot \bold r}$ one obtains a term proportional to $\bold \sigma \cdot
\bold k \, e^{- i \, \bold k \cdot \bold r}$ which can be absorbed in $g$.
To eq.(\ref{op}), one
must also add the contribution from antiquarks
\be  \label{opp}
{\cal H}''=X^{M''}_{\bar q_j \bar q'_{j}} \,
  \Bigl[ \, -g \, (\bold \sigma_j \cdot \bold k)
\, e^{- i \, \bold k \cdot \bold r_j} \, + \, h \,
(\bold \sigma_j \cdot \bold p_j ) \, e^{- i \, \bold k
 \cdot \bold r_j} \, \Bigr] \, .
\ee
All two--body decays are thus given in terms of two parameters $h$ and $g$.
The form (\ref{op}) is common to both the elementary--emission model
and the pair creation model of strong meson decays~\cite{LeY}, the only
difference being the values of the coefficients $g$ and $h$. We assume that
these decay constants are flavor independent. Although our results can be
easily generalized to the case in which the decay constants of
$u$, $d$ and $s$ are different, we prefer, for the purpose of the
present article, to keep the transition operator in its simplest possible form.
All flavor dependence is thus in the matrices $X$.
\par
The decay widths are computed from
\be \label{ampl}
\Gamma(M \longrightarrow M' + M'' )= {k \over {2 \pi}}\,
{{E_{M'}} \over {E_M}} \, \vert \bra M' \vert {\cal H} \vert M \ket \vert^2\, ,
\ee
where ${\cal H}={\cal H'}+{\cal H''}$ and
$E_{M'}$ and $E_M$ are the total energy of the final meson and the mass
of the initial meson. The ratio ${{E_{M'}} \over {E_M}}$ can be rewritten
in a frame independent way as a function of the masses of mesons appearing
in eq.(\ref{dec}):
\be \label{chi}
\chi={{E_{M'}} \over {E_M}}={1\over 2}+{{m_{M'}^2-m_{M''}^2}\over {2\,
m_{M}^2}}
\, .
\ee
$M''$ is taken to be the lighter of the two mesons
in the final state. If the two mesons $M'$ and $M''$ are identical,
an extra factor of ${1 \over 2}$ is included. The operators (\ref{op}) and
(\ref{opp}) are expressed in the quark basis $\vert q_i \bar q_j \ket$,
but can be easily reexpressed in the $SU(3)_f$ basis by
writing $X$ in $SU(3)_f$ tensorial notation.
In either case, while in the evaluation of the electromagnetic decays~\cite{ID}
the flavor part was trivial and all the complication was in the space--spin
part, in the case of strong decays both parts require heavy algebra.
Explicit expressions for the matrix $X$ in the quark basis are given
in Ref.~\cite{GI}, and they can be used directly for the evaluation of
the matrix elements in (\ref{ampl}).
If instead $SU_f(3)$ wave functions are used, the evaluation of the matrix
elements in  (\ref{ampl}) requires the knowledge of the $SU_f(3)$ and
$SU_I(2)$ isoscalar factors
\be
\left( \matrix{ M' & M'' \cr
                I_1 Y_1 & I_2 Y_2 } \right \vert
\left. \matrix{ M \cr I Y } \right)
\left( \matrix{ I_1 & I_2 \cr
                I_{z1} & I_{z2} } \right\vert
\left. \matrix{I \cr I_z } \right)
\, .
\ee
These are given in Ref.~\cite{JJS}. \par
We have used both methods to evaluate the 25 strong decay widths
shown in Table II. The use of both methods provides us with a check
of the correctness of the results (the two methods differ only by
an overall normalization factor). The computed widths are written in terms of
some kinematic factors, the spin--flavor part and the form factors
$F_i(\nu,k)$,
 $i=1,\ldots,5$. The form factors $F_i$ contain all the information of the
hadronic structure: they depend on the momentum of the emitted pseudoscalar
meson $k$ and on the ratio  $\nu=m_{q(\bar q)}/(m_q+m_{\bar q})$,
where $m_q$ and $m_{\bar q}$ are the quark and antiquark masses of the decaying
meson. This dependence naturally arises in the calculation
of the radial part of the matrix elements.
The expressions in Table II include the sum and averaging over
the components of the final and initial states, respectively.
The calculations of the space part of the matrix elements,
$e^{- i \, \bold k \cdot \bold r}$ for the first term and
$\bold p \, e^{- i \, \bold k \cdot \bold r}$ for the second term,
are done using the algebraic method discussed in Sect.II of Ref.~\cite{ID},
which is not repeated here. The relevant matrix elements are tabulated
in Appendix A of Ref.~\cite{ID}. \par

\subsection*{C. Form factors}
The widths in Table II are given in terms of form factors $F_i(k)$. By using
algebraic methods, as discussed in Ref.~\cite{ID}, these form factors can be
evaluated in closed form in three different situations:
\begin{description}
\item[(I) $U(3)$:] the $U(3)$ form factors are given in Table IIIa. They
are all combinations of exponentials $\exp(-\alpha k^2)$,
and polynomials in $k^2$. These form factors are the form factors
of the non--relativistic harmonic oscillator quark model or variations
of it~\cite{QM,GI}.
\item[(II) $SO(4)$:] the $SO(4)$ form factors are given in Table IIIb.
They are combinations of spherical Bessel functions. These are the form factors
of a rigid string with quarks sitting at its ends.
\item[ (III) $SO(4)^*$:] a more realistic case is that in which
the meson is emitted from a string whose length is given by the
probability distribution $\beta \exp(-2 \beta/a)$. This
produces form factors with a power law behaviour for large $k$.
The corresponding expressions are
obtained by replacing, in the $SO(4)$ form factors, the spherical Bessel
functions $j_l(\beta k \nu)$ by
\be
\tilde j_l(ak\nu)=
{{\int^\infty_0 d\beta (\beta e^{-2\beta/a})j_l(\beta k \nu)}
\over
{\int^\infty_0 d\beta (\beta e^{-2\beta/a})}    } \, .
\ee
The integrated functions $\tilde j_l$ are shown in Table IIIc.
This situation is denoted here by the $SO(4)^*$.
\end{description}
The three situations (I), (II), and (III) represent three extreme
situations which encompass a large variety of hadronic structure models.
The form factors of each situations contain one parameter
($\alpha$, $\beta$ or $a$) characterizing the average size of the mesons.

\section{Analysis of Experimental Data}

The results of Table II can be used to analyze the experimental data.
A least square fit to the available data
gives the values of the parameters shown
in Table IV. With these values one can calculate
the strong decay widths shown
in Table V.
When several final charge states are possible, the appropriate isospin
factors have been included going from Table II to Table V.
Also, in calculating the values of Table V the following mixing angles
have been used
\be
\theta_P=-23^\circ, \qquad \theta_V=38^\circ, \qquad \theta_T=26^\circ\, .
\ee
These values have been kept fixed. The value of the pseudoscalar mixing
angle $\theta_P=-23^\circ$ is consistent with that determined in Ref.~\cite{IM}
from a fit to the meson masses. The value of $\theta_V$, when converted
to the notation of Ref.~\cite{ID}, is $2.7^\circ$, which is somewhat different
form the value $4.3^\circ$ of this reference. Finally, the value of $\theta_T$
for the tensor mesons is the same as reported in \cite{PDB}.\par
In order to display clearly the situation, we show in Figs.2--6 a comparison
between the experimental data and the calculations. One can see that
the agreement between calculated values and experiment is
in most cases good, with few exceptions discussed in subsections A and B
below. One can also
see that, within the range of momenta tested by the decays of Table II,
the three classes of form factors produce similar results.
This is part of the reason why the harmonic oscillator
quark model, although in clear contradiction with experiment for large $k^2$,
provides a reasonably good description of known decay widths.\par
In view of the fact that the calculated values agree with the experimental
values on the average within a factor of two or better, one can address
specific problems related to meson spectroscopy.

\subsection*{A. Decays of $K_1^+$ and the tensor force}

Table V and Figs.2 and 6 show that two of the calculated decays which
are in bad agreement with experiment are:
\bea \label{mix}
K_1^+(1270) & \longrightarrow & \rho^0 K^+ \, ,\cr
K_1^+(1400) & \longrightarrow & \rho^0 K^+  \, .
\eea
The quantum numbers of $K_1^+(1270)$
and $K_1^+(1400)$ are $^1\! P_1 $ and $ ^3\! P_1 $ respectively.
The calculated decay widths can be brought into better agreement
with experiment by mixing of the two states,
\bea  \label{angle}
\vert K_1^+(1270)\ket & = & \cos \varphi \, \vert ^1\! P_1 \ket
                       + \sin \varphi \, \vert  ^3\! P_1 \ket \, ,\cr
\vert K_1^+(1400)\ket & = & -\sin \varphi \, \vert ^1\! P_1 \ket
                       + \cos \varphi \, \vert  ^3\! P_1 \ket \, .
\eea
The mixing between $^1\! P_1 $ and $ ^3\! P_1 $ can be produced only by
a tensor force. The decays (\ref{mix}) provide therefore further evidence
for the occurrence of a tensor force, similar to the evidence obtained from
baryons (mixing of $N(1535)$ and $N(1650)$).
Since one is looking for clues for the correctness of QCD in the
non--perturbative regime, the decays (\ref{mix}) appear to confirm the
occurrence of a tensor force
\be
S_{12}=A \, \bigl[ T_s^{(2)} \otimes  T_{\cal R}^{(2)} \bigr] \, ,
\ee
where $T_s^{(2)}$ is an operator of rank 2 acting on the spin variables and
$ T_{\cal R}^{(2)}$ an operator of rank 2 acting on the space variables.
The tensor force can be formally derived in QCD from one--gluon exchange.
\par
The value $\varphi=46.8^\circ$ gives
\bea
\Gamma(K_1(1270)) \to \rho K) & = 15.3 \ {\rm MeV} \, , \cr
\Gamma(K_1(1400)) \to \rho K) & = 5.2  \ {\rm MeV} \, ,
\eea
still not in complete agreement with experiment ($37.8\pm 13.8$ and
$5.2\pm 5.2$, respectively), but in much better agreement
than the unmixed values. The particular value of $\varphi$ has been
computed using the results of the $U(3)$ fit, but similar results
hold in the $SO(4)$ and $SO(4)^*$ symmetry.

\subsection*{B. Decays of $f_2(1270)$, $f'_2(1525)$ and glueballs}

Table V shows that the only other calculated decay width which is in bad
agreement with experiments is
\be
f'_2(1525) \longrightarrow \pi \pi \, .
\ee
The decays of $f_2$ and $f'_2$ into $K\bar K$, $\eta \eta$, $\pi \pi$
can be used to study admixtures of glueballs components into
$q \bar q$ states. We shall readdress this question in a later publication
where the mixing of $f_2(1270)$, $f'_2(1525)$
with the glueball candidate $f_2(1720)$ will be analyzed in great detail.
Here we only stress that the anomalously large deviation
$\Gamma(f'_2 \to \pi \pi)_{calc} / \Gamma(f'_2 \to \pi \pi)_{exp} \simeq 10$
seems to indicate that other components ($q^2 \bar q^2$ or gluonic)
are present in the region of $f_2(1270)$
(the calculated width of the $f'_2 \to \pi \pi$ decay could be brought into
better agreement with experiment
by using a different value for the tensor mixing angle,
$\theta_T\simeq 32.7^\circ$, but this mixing angle would produce a bad
agreement
of other $f_2$ and $f'_2$ decays, expecially for $f_2 \to \eta\,\eta$).

\subsection*{C. Decays of $\phi(1020)$ and tests of kinematics}

The two decays
\bea
\phi & \longrightarrow & K^+ \, K^- \cr
\phi & \longrightarrow & K^0_L \, K^0_S
\eea
provide a test of the kinematics used in the calculation.
The matrix elements for the two processes are identical (and for this reason
the decay $\phi  \to K^0_L \, K^0_S$ is not shown in Tables II and V),
while the values
of $k$ are slightly different, $k=127$ MeV/c and $k=110$ MeV/c respectively.
The experimental ratio
\be \label{ratio}
{{\Gamma(\phi  \longrightarrow  K^+ \, K^-)}
\over
 {\Gamma(\phi  \longrightarrow  K^0_L \, K^0_S )}}
= (1.43 \pm 0.06)
\ee
can be compared with that calculated in (I), (II) and (III):
\be
1.534 \quad (I)\,,  \  \qquad   1.535 \quad (II)\,,
 \ \qquad 1.531 \quad (III)\, . \  \qquad
\ee
The agreement is good, and practically the same in all three situations.
This is an important point, because for decays of relativistic particles it is
not at all obvious what kinematic factors should be used in connection
with the transition operator ${\cal H}$ of Eqs.(\ref{op})-(\ref{opp}).\par
We note in passing that, in the early
days of hadronic spectroscopy \cite{Glas}, the widths were calculated
by introducing a form factor $\vert k^2 / (k^2 + R^{-2})\vert^l$
and a kinematic factor $k/M$, i.e.
\be
\Gamma \propto \Bigl\vert {k^2 \over {k^2 + R^{-2} }} \Bigr\vert^l
{k \over M}
\, ,
\ee
where $M$ is the mass of the decaying meson, $R^{-1}$ a
measure of the size of the system ($R^{-1}=350$ MeV/c)
and $l$ is the angular momentum of the decaying particle.
The kinematic factor and form factor give, for the
ratio (\ref{ratio}), $1.52$, also in agreement with experiment.

\subsection*{D. Decays of $\omega(783)$ and isospin mixing}

Although not directly related to the results of Table V,
we comment briefly on another problem of interest which can be studied
with strong decays: isospin mixing. The decay $\omega \to \pi \pi$
(not shown in Table V) is forbidden by G--parity.
However, a small mixing of $\omega^0$ with $\rho^0$ will allow the decay to go.
{}From the observed decay width
\be
\Gamma(\omega^0 \longrightarrow \pi^+ \pi^-)= (186.3 \pm 27.5) \ {\rm keV}\, ,
\ee
one finds
\bea
\vert \rho^0 \ket & = &
                       \cos 2.0^\circ \, \vert I=1 \ket + \sin 2.0^\circ \,
                                  \vert I=0  \ket \, ,\cr
\vert \omega^0 \ket & = & -\sin 2.0^\circ \, \vert I=1 \ket + \cos 2.0^\circ
\, \vert I=0 \ket \, .
\eea
The value of the mixing angle $ 2.0^\circ $
is somewhat smaller than the value $6.8^\circ$ determined from the analysis of
the radiative decays $\rho^0 \longrightarrow \gamma \, \pi^0$,
$\rho^\pm \longrightarrow \gamma \, \pi^\pm$ and $\omega^0 \longrightarrow
\gamma \, \pi^0$~\cite{ID}.

\section{Predictions}

The parameterization of the strong decay widths described here allows one to
make
predictions for unknown widths within a factor of 2 or better.
These predictions can in turn be used to extract ``new" physics.\par

\subsection*{A. Decays of $a_0(^3\!P_0)$ and missing states}

As an example we address here two problems: the so--called ``problem
of missing states" and the nature of $a_0(980)$. We use the method of
Sect. 2 to calculate the decay matrix of $a_0(^3\!P_0)$, $a_1(^3\!P_1)$ and
$a_2(^3\!P_2)$ into $\rho \, \pi$, $\eta \, \pi$, $K \bar K$.
The allowed decays of $a_2$ and $a_1$
are already given in Table II.
The decays of $a_0$ are shown in Table VI.
The predicted decay matrix is shown in Table VII, where we report the
results of the $U(3)$ symmetry (the $SO(4)$ and $SO(4)^*$ situations
have similar qualitative behaviour). We note that the decays
which are experimentally seen are in excellent agreement with calculations,
and that the fact that the transition operator of Sect. 2 gives
\be
\Gamma(a_1(1260) \longrightarrow \eta \, \pi) \, = \, 0 \qquad {\rm and} \qquad
\Gamma(a_1(1260) \longrightarrow K \, \bar K) \, = \, 0 \, ,
\ee
is in agreement with the non--observation of these decays. Most importantly,
we find that the decays of the $a_0$ give a total width of about 420--940 MeV,
for $a_0$ masses in the range 1200--1400 MeV
with two branches $\eta \, \pi$ and $K \bar K$ (since $\Gamma(a_0(^3\!P_0)
\to \rho \pi)=0$ with the transition operator of Sect.2).
This result suggests strongly that the $a_0(980)$, which has a total decay
width of $57\pm 11$ MeV, is {\it not} the $q \bar q$ state $^3\!P_0$.
Furthermore, the width of the
$a_0(^3\!P_0)$ state is so large that it might escape detection. This
indeed may be one of the reason why some states
predicted by models of hadron structure are either not seen
or marginally seen (missing states). The combination of the result of
Ref.~\cite{IM}, which predicts the state $a_0(^3\!P_0)$ at 1273 MeV and the
present paper, strongly suggest that the state $a_0(1320)$ reported at the
Hadron 89 Conference and omitted from the Summary Tables of Ref.~\cite{PDB}
is a good candidate for the missing $a_0(^3\!P_0)$ state.

\section{Conclusions}

We have presented here a reanalysis of the strong decays of $q\bar q$ mesons
similar in spirit to the earlier analyses in terms of flavor $SU_f(3)$
symmetry~\cite{Glas}, but where symmetries are also used to deal
with the space part of the hadron wavefunctions.
In particular, for mesons, we have used the spectrum generating algebra
$ {\cal G}=U(4) \otimes SU_s(2) \otimes SU_f(3) \otimes SU_c(3)$,
and considered two branchings of $U(4)$, into $U(3)$ and $SO(4)$.
The use of $U(4)$ provides us with explicit expressions for the form
factors, and thus allows us to compute analytically {\it all} decay widths.
\par
We note that the results of Table V show that strong decay widths
are not much dependent on models of hadronic structure ($U(3)$,
$SO(4)$ and $SO(4)^*$) but they depend almost exclusively on the spin--flavor
part of the meson wave functions. The parameterization of the decay operator
(\ref{op}) and (\ref{opp}) appears to give all decay widths within a factor
of 2. \par
Having constructed the formalism for masses~\cite{IM},
electromagnetic~\cite{ID}, weak~\cite{IL} and strong decays, we are now
in a position to analyze any number of Ref.~\cite{PDB}, related to these
quantities. Deviations from symmetry parameterizations can be used to extract
``new" physics. Particularly important in meson spectroscopy
are the searches for gluonic components and multiquark configurations
($q^2 \bar q^2$, $\ldots$). We believe we have now a method in which this
search
can be done in a somewhat quantitative way.
We also find it important to test the accuracy of the predictions
of this article and of Ref.~\cite{ID}, by doing new experiments.
New hadronic facilities, such as the $\phi$ factory, presently under
construction at Frascati, Italy, may help in this respect, expecially in the
study of the decays of $a_0$.
Values of the calculated decay widths for any two--body strong decay
can be obtained from us upon request.

\section*{Acknowledgement}
This work was supported in part under D.O.E. Grant no. DE-FG02-91ER40608.
One of us (CG) was also partially supported by {\it Fondazione Della Riccia}.
\par
\vspace{2.0cm}
\noindent
$^*$ also {\it Istituto Nazionale di Fisica Nucleare}, Sezione di Pavia, Italy.

\newpage
\section*{Figure and Table Captions}
\begin{description}
\item[Fig.1] Elementary meson emission: part (a) refers to the
operator ${\cal H}''$, part (b) to ${\cal H}'$.
\item[Figs.2] Comparison between calculated and experimental values of the
form factor $\vert F_1(k)\vert$ as a function of $k$. The experimental
points are the values of $\sqrt{\Gamma_{exp}}$ divided by the appropriate
factors appearing in Table II. The theoretical curves are the form factors
$\vert F_1(1/2,k)\vert$ with the values of $g$ and $h$ (or $h'$)
given in Table IV
(full line, $U(3)$; dashed, $SO(4)$; dotted, $SO(4)^*$).
\item[Fig.3] Same as Fig.2, but for $\vert F_2(k)\vert$. The experimental
value at $k=$0.36 which is in bad agreement with calculations is that
of $K_1(1270) \to \rho K$ discussed in Sect.3.A.
\item[Fig.4] Same as Fig.2, but for $\vert F_3(k)\vert$.
The experimental value at $k=$1.52 (in bad agreement with predictions)
is that of  $K_1(1400) \to \rho K$ discussed in Sect.3.A.
\item[Fig.5] Same as Fig.2, but for $\vert F_4(k)\vert$.
The experimental value at $k=$3.15 is that of $f'_2 (1525) \to \pi\pi$
discussed in Sect.3.B.
\item[Fig.6] Same as Fig.2, but for $\vert F_5(k)\vert$.
\item []
\item[Table I] Quantum numbers assignments of $q \bar q$
states in $U(3)$ and $SO(4)$.
\item[Table II] Analytic strong decay widths for selected light mesons.
The pseudoscalar, vector and tensor mixing angles are denoted by
$\theta_P$, $\theta_V$ and $\theta_T$.
$m$ indicates the $u\!-\!d$ mass ($\sim 250$ MeV),
$m_s$ the strange quark mass ($\sim 400$ MeV) and
$\chi$ is defined in eq.(\ref{chi}).
\item[Table III] Form factors appearing in Table II. The parameter $h'$
in (b) and (c) is equal to $h\zeta$, where $\zeta$ is the scale of
the momenta~\cite{ID}. The phase conventions for the $SO(4)$
wave functions is the same as for $U(3)$.
\item[Table IV] Values of the best fit parameters. $g$ and $h$ are in fm,
$h'$ is dimensionless.
\item[Table V] Comparison between the experimental \cite{PDB} and calculated
decay widths. All widths are in MeV, $k$ is in ${\rm fm}^{-1}$.
\item[Table VI] Analytic strong decay widths for $a_0(^3\!P_0)$  decays.
\item[Table VII] Calculated strong decays of $a_0$, $a_1$ and $a_2$ mesons.
Since the mass of $a_0$ is not known, the values in the table are for
$m_{a_0}$=1200--1400 MeV. All values are in MeV.

\end{description}
\newpage

\begin{center}
{\bf Table I}
\end{center}
\begin{center}
\begin{tabular}{|ll|ccc|ccc|}
\hline \hline
& &  \multicolumn{3}{|l}{$U(3)$ symmetry} &
\multicolumn{3}{|l|}{$SO(4)$ symmetry} \\
Meson & $J^{PC}$ & $n$ & $L$ & $S$ & $v$ & $L$ & $S$ \\ \hline
\multicolumn{8}{c}{$\pi$ family} \\
\hline
$\pi$ & $0^{-+}$ &            0 & 0 & 0 &  0 & 0 & 0 \\
$\rho(770)$ & $1^{--}$ &      0 & 0 & 1 &  0 & 0 & 1 \\
$b_1(1235)$ & $1^{+-}$ &      1 & 1 & 0 &  0 & 1 & 0 \\
$a_1(1260)$ & $1^{++}$ &      1 & 1 & 1 &  0 & 1 & 1 \\
$a_2(1320)$ & $2^{++}$ &      1 & 1 & 1 &  0 & 1 & 1 \\
$\rho(1450)$ & $1^{--}$ &     2 & 0 & 1 &  1 & 0 & 1 \\
$\pi_2(1670)$ & $2^{-+}$ &    2 & 2 & 0 &  0 & 2 & 0 \\
$\rho_3(1690)$ & $3^{--}$ &   2 & 2 & 1 &  0 & 2 & 1 \\
$\rho(1700)$ & $1^{--}$ &     2 & 2 & 1 &  0 & 2 & 1 \\ \hline
\multicolumn{8}{c}{$K$ family} \\
\hline
$K$ & $0^-$ &                 0 & 0 & 0 &  0 & 0 & 0 \\
$K^*(892)$ & $1^-$ &          0 & 0 & 1 &  0 & 0 & 1 \\
$K_1(1270)$ & $1^+$ &         1 & 1 & 0 &  0 & 1 & 0 \\
$K^*(1370)$ & $1^-$ &         2 & 0 & 1 &  1 & 0 & 1 \\
$K_1(1400)$ & $1^+$ &         1 & 1 & 1 &  0 & 1 & 1 \\
$K^*_0(1430)$ & $0^+$ &       1 & 1 & 1 &  0 & 1 & 1 \\
$K^*_2(1430)$ &  $2^+$ &      1 & 1 & 1 &  0 & 1 & 1 \\
$K^*(1680)$ & $1^-$ &         2 & 2 & 1 &  0 & 2 & 1 \\
$K_2(1770)$ & $2^-$ &         2 & 2 & 1 &  0 & 2 & 1 \\
$K^*_3(1780)$ & $3^-$ &       2 & 2 & 1 &  0 & 2 & 1 \\
$K^*_4(2045)$ & $4^+$ &       3 & 3 & 1 &  0 & 3 & 1 \\ \hline
\end{tabular}
\end{center}
\newpage
\begin{center} {\bf Table I (continued)}
\end{center}
\begin{center}
\begin{tabular}{|ll|ccc|ccc|}
\hline \hline
& &  \multicolumn{3}{|l}{$U(3)$ symmetry} &
\multicolumn{3}{|l|}{$SO(4)$ symmetry} \\
Meson & $J^{PC}$ & $n$ & $L$ & $S$ & $v$ & $L$ & $S$ \\ \hline
\multicolumn{8}{c}{$\eta$ family} \\
\hline
$\eta$ & $0^{-+}$ &           0 & 0 & 0 &  0 & 0 & 0 \\
$\eta'(958)$ & $0^{-+}$ &     0 & 0 & 0 &  0 & 0 & 0 \\
$\omega(783)$ & $1^{--}$ &    0 & 0 & 1 &  0 & 0 & 1 \\
$\phi(1020)$ & $1^{--}$ &     0 & 0 & 1 &  0 & 0 & 1 \\
$h_1(1170)$ & $1^{+-}$ &      1 & 1 & 0 &  0 & 1 & 0 \\
$f_2(1270)$ & $2^{++}$ &      1 & 1 & 1 &  0 & 1 & 1 \\
$f_1(1285)$ & $1^{++}$ &      1 & 1 & 1 &  0 & 1 & 1 \\
$\eta(1295)$ & $0^{-+}$ &     2 & 0 & 0 &  1 & 0 & 0 \\
$\omega(1390)$ & $1^{--}$ &   2 & 0 & 1 &  1 & 0 & 1 \\
$f_1(1510)$ &  $1^{++}$ &     1 & 1 & 1 &  0 & 1 & 1 \\
$f'_2(1525)$ & $2^{++}$ &     1 & 1 & 1 &  0 & 1 & 1 \\
$\omega(1600)$ &  $1^{--}$ &  2 & 2 & 1 &  0 & 2 & 1 \\
$\omega_3(1600)$ & $3^{--}$ & 2 & 2 & 1 &  0 & 2 & 1 \\
$\phi(1680)$ &   $1^{--}$ &   2 & 0 & 1 &  1 & 0 & 1 \\
$\phi_3(1850)$ & $3^{--}$ &   2 & 2 & 1 &  0 & 2 & 1 \\
$f_4(2050)$ &   $4^{++}$ &    3 & 3 & 1 &  0 & 3 & 1 \\  \hline
\end{tabular}
\end{center}
 \newpage
\begin{center}
{\bf Table II}
\end{center}
\begin{center}
\begin{tabular}{ll} \hline \hline
\\
\multicolumn{2}{c}{$^3\!S_1 \longrightarrow ^1\!S_0$}\\
$\Gamma_{\rho^0(770) \to \pi^- \pi^+}$
                   &  $= {k \over {16 \pi}} \, \vert F_1(1/2,k) \vert^2  $ \\
$\Gamma_{\phi(1020) \to K^-K^+}$ & $=
                    {{3k} \over { 64 \pi}} \, \cos^2 \theta_V \,
                          \vert F_1(1/2,k) \vert^2  $\\
$\Gamma_{K^{*0}(892) \to \pi^-K^+}$ & $=
      {k \over {16 \pi}} \, \chi \, \vert F_1({m \over { m+m_s}},k) \vert^2
$\\
\\
\multicolumn{2}{c}{$^3\!S_1 \longrightarrow ^3\!S_1$}  \\
$\Gamma_{\phi(1020) \to \rho^- \pi^+}$ & $ =
        {k \over {6 \pi}} \, \chi \, (\sin\theta_V -
                {{\cos\theta_V} \over \sqrt{2}})^2 \,
                   \vert F_1(1/2,k) \vert^2  $\\ \\
\multicolumn{2}{c}{$^1\!P_1 \longrightarrow ^3\!S_1$}\\
$\Gamma_{b_1^+(1235) \to \omega \, \pi^+}$ & $=
               {k \over {12 \pi}} \, \chi \,
({{\sin\theta_V} \over \sqrt{2}}+\cos\theta_V )^2 \,
                       \vert F_2(1/2,k) \vert^2  $\\
$\Gamma_{K_1^+(1270) \to \rho^0 K^+}$ & $= {k \over {32 \pi}}\,  \chi \,
                \vert F_2({m \over { m+m_s}},k) \vert^2  $\\
$\Gamma_{K_1^+(1270) \to K^{0*} \pi^+}$ & $= {k \over {16 \pi}}\,  \chi \,
                \vert F_2({m_s \over { m+m_s}},k) \vert^2  $\\
\\
\multicolumn{2}{c}{$^3\!P_1 \longrightarrow ^3\!S_1$}    \\
$\Gamma_{a_1^0(1260) \to  \rho^- \pi^+}$ & $=   {k \over {8 \pi}}\,  \chi \,
               \vert F_3(1/2,k) \vert^2  $\\
$\Gamma_{K_1^+(1400) \to K^{0*} \pi^+}$ & $ = {k \over {16 \pi}}\,  \chi \,
               \vert F_3({m_s \over { m+m_s}},k) \vert^2  $\\
$\Gamma_{K_1^+(1400) \to \rho^0 K^+}$ & $ = {k \over {32 \pi}} \, \chi \,
               \vert F_3({m \over { m+m_s}},k) \vert^2  $\\
\\
\hline
\end{tabular}
\end{center}
 \newpage
\begin{center}
{\bf Table II (continued)}
\end{center}
\begin{center}
\begin{tabular}{ll} \hline \hline
\\
\multicolumn{2}{c}{$^3\!P_2 \longrightarrow ^1\!S_0$}      \\
$\Gamma_{a_2^+(1320) \to \eta \pi^+}$ & $ = {k \over {20 \pi}}\,  \chi\,
                     (\sin\theta_P - {{\cos\theta_P} \over \sqrt{2}})^2 \,
                       \vert F_4(1/2,k) \vert^2  $\\
$\Gamma_{a_2^0(1320) \to  K^-K^+}$ & $= {{3k} \over { 320 \pi}}
                    \, \vert F_4(1/2,k) \vert^2  $\\
$\Gamma_{K_2^{*0}(1430) \to \pi^-K^+}$ & $ = {{3k} \over { 80 \pi}} \,  \chi \,
                      \vert F_4({m \over { m+m_s}},k) \vert^2  $\\
$\Gamma_{K_2^{*0}(1430) \to \eta K^0}$ & $ ={k \over {160 \pi}} \, \chi \,
                     \vert (\sin \theta_P+\sqrt{2} \cos \theta_P) \,
                      F_4({m_s \over { m+m_s}},k) \, +\,
                           (\sin\theta_P- {{\cos\theta_P} \over \sqrt{2}})\,
                      F_4({m \over { m+m_s}},k) \vert^2  $\\
$\Gamma_{f'_2(1525) \to  K^-K^+}$ &  $=  {k \over {160 \pi}}\,
                        ({{\cos\theta_T} \over \sqrt{2}}+ 2 \sin\theta_T)^2\,
                   \vert F_4(1/2,k) \vert^2  $\\
$\Gamma_{f'_2(1525) \to  \eta \eta }$ & $ = {1\over 2}\,  {k \over {180 \pi}}\,
              \vert \Bigl\{ 4\,  (\sin\theta_P
                 -{{\cos\theta_P} \over \sqrt{2}})\,
              (\sin\theta_T -{{\cos\theta_T} \over \sqrt{2}}) \,
              (-\sin\theta_P - {{\cos\theta_P} \over 4})\, + \, $\\
            & $ 2 \, (\sin \theta_P+\sqrt{2} \cos \theta_P) \,
               (\sin \theta_T+\sqrt{2} \cos \theta_T)  \,
           ( {{ \cos \theta_P}\over 2} -\sin\theta_P) \Bigr\} \, F_4(1/2,k)
                         \vert^2  $\\
$\Gamma_{f'_2(1525) \to \pi^- \pi^+}$ & $= {k \over {40 \pi}}\,
                    (\sin\theta_T -{{\cos\theta_T} \over \sqrt{2}})^2\,
                       \vert F_4(1/2,k) \vert^2  $\\
$\Gamma_{f_2(1270) \to \pi^- \pi^+}$ & $= {k \over {40 \pi}} \,
                   ({{\sin\theta_T} \over \sqrt{2}}+\cos\theta_T)^2\,
                  \vert F_4(1/2,k) \vert^2  $\\
$\Gamma_{f_2(1270) \to K^-K^+}$ & $ = {k \over {160 \pi}} \,
                   ({{\sin\theta_T} \over \sqrt{2}}-2\cos\theta_T)^2\,
                     \vert F_4(1/2,k) \vert^2  $\\
$\Gamma_{f_2(1270) \to \eta \eta }$ & $ = {1\over 2}\,   {k \over {180 \pi}} \,
             \vert \Bigl\{ 4\,({{\sin\theta_P}
                    \over \sqrt{2}}+\cos\theta_P) \,
       ({{\sin\theta_T} \over \sqrt{2}}+\cos\theta_T) \,
       (-\sin\theta_P - {{\cos\theta_P} \over 4})\, +\, $ \\
        & $ 2 (\sqrt{2}\sin \theta_P-\cos\theta_P)\,
         (\sqrt{2}\sin \theta_T-\cos\theta_T) ({{\cos\theta_P} \over 2}
 -\sin\theta_P) \Bigr\} \,     F_4(1/2,k) \vert^2  $\\
\\
\multicolumn{2}{c}{$^3\!P_0 \longrightarrow ^1\!S_0$}        \\
$\Gamma_{K_0^{*0}(1430) \to \pi^-K^+}$ & $ = {{3k} \over { 16 \pi}}\,   \chi \,
                  \vert F_5({m \over { m+m_s}},k) \vert^2  $\\
\\
\multicolumn{2}{c}{$^3\!P_2 \longrightarrow ^3\!S_1$}          \\
$\Gamma_{a_2^+(1320) \to \rho^0  \pi^+}$ & $ = {{3k} \over { 40 \pi}} \,  \chi
\,
                     {3 \over 2}\,
                    \vert F_4(1/2,k) \vert^2  $\\
$\Gamma_{K_2^{*0}(1430) \to  K^{*+} \pi^-}$ & $= {{3k} \over { 80 \pi}} \,
\chi \,
                                  {3 \over 2}\,
                   \vert F_4({m_s \over { m+m_s}},k)\vert^2  $\\
$\Gamma_{K_2^{*0}(1430) \to  \rho^- K^+}$ & $= {{3k} \over { 80 \pi}} \,
\chi\,
                         {3 \over 2} \,
                    \vert F_4({m \over { m+m_s}},k)\vert^2  $\\
$\Gamma_{K_2^{*0}(1430) \to \omega K^0}$ & $= {k \over { 80 \pi}}\,  \chi \,
                           {3 \over 2} \,
    \vert (\sqrt{2}\sin \theta_V-\cos\theta_V) \,
               F_4({m_s \over { m+m_s}},k)\,
   + \, ({{\sin\theta_V} \over \sqrt{2}}+\cos\theta_V)\,
            F_4({m \over { m+m_s}},k)\vert^2  $\\
\\
\hline
\end{tabular}
\end{center}
\newpage
\begin{center}
{\bf Table III}\\
\vspace{1.0cm}
(a) $U(3)$ symmetry
\end{center}
\begin{center}
\begin{tabular}{l}
\hline \hline \\
$F_1(\nu,k)=\Bigl[g+{\nu \over 2} h\Bigr] \, k \,
            \exp(-{{\alpha^2\nu^2 k^2}\over {4 }}) $\\  \\
$F_2(\nu,k)={i \over \sqrt{2}} \Bigl\{ \bigl\vert {h \over \alpha} -
{{k^2 \nu}\alpha} \Bigl[g+{\nu \over 2} h\Bigr] \bigr\vert^2
            + {{2h^2}\over \alpha^2} \Bigr\}^{1/2} \,
\exp(-{{\alpha^2 \nu^2 k^2}\over {4}}) $\\  \\
$F_3(\nu,k)=i \Bigl\{{h^2\over \alpha^2} +2 \bigl\vert
   -{ h\over \alpha} + {1\over 2 }{{k^2 \nu}\alpha}
\Bigl[g+{\nu \over 2} h\Bigr]  \bigr\vert^2\}^{1/2} \,
 \exp(-{{\alpha^2 \nu^2 k^2}\over {4 }}) $\\   \\
$F_4(\nu,k)={i \over \sqrt{3}}{{k^2 \nu} \alpha}
\Bigl[g+{\nu \over 2} h\Bigr]\, \exp(-{{\alpha^2
                              \nu^2 k^2}\over {4 }}) $\\  \\
$F_5(\nu,k)={i \over \sqrt{6}}\, \Bigl\{ {{k^2 \nu} \alpha} \Bigl[
   g+{\nu \over 2} h\Bigr]  -{3h \over \alpha} \Bigr\} \,
   \exp(-{{\alpha^2 \nu^2 k^2}\over {4 }}) $\\     \\
\hline
\end{tabular}
\end{center}
\begin{center}
\vspace{1.0cm}
(b) $SO(4)$ symmetry
\end{center}
\begin{center}
\begin{tabular}{l}
\hline \hline      \\
$F_1(\nu,k)=g\, k\, j_0(k\beta \nu)\, +\, h'  \,  j_1(k \beta \nu) $ \\ \\
$F_2(\nu,k)=i \Bigl\{ \vert \, g\, k\,  \sqrt{3} \,
j_1(k \beta \nu)\, - \, {{h'} \over {\sqrt{3}}}\,
[j_0(k \beta \nu)\, -\, 2\, j_2(k \beta \nu)\, ]\vert^2
\, +\, {2\over 3}\, {h'^2}\,
 \vert j_0(k \beta \nu)+j_2(k \beta \nu)\vert^2 \Bigr\}^{1/2} $\\ \\
$F_3(\nu,k)=i \Bigl\{ {2\over 3}\,{h'^2} \, \vert j_0(k \beta \nu)
+j_2(k \beta \nu)\vert^2
+\vert \, g\, k\,  \sqrt{3}\, j_1(k \beta \nu)\, -\,  {{h'} \over
{\sqrt{3}}}
[2\, j_0(k \beta \nu)-j_2(k \beta \nu)] \vert^2 \Bigr\}^{1/2} $\\   \\
$F_4(\nu,k)=i\, \sqrt{2}\,  [-g\, k\,
j_1(k \beta \nu)-\, h' \, j_2(k \beta \nu)]$ \\  \\
$F_5(\nu,k)=i \, [g\, k\,  j_1(k \beta \nu)-
 {h'}\, j_0(k \beta \nu)]  $ \\  \\
\hline
\end{tabular}
\end{center}
\vspace{2.0cm}
\begin{center}
\vspace{1.0cm}
(c) $SO(4)^*$ symmetry: replace $j_l$ in part (b) by $\tilde j_l$.
\end{center}
\begin{center}
\begin{tabular}{l}
\hline \hline    \\
$ \tilde j_0(ka \nu)={1 \over {1 +Q^2}} $ \\
\\
$ \tilde j_1(ka \nu)={1 \over {Q(1 +Q^2)}} \{ [Q+{1\over Q}] {\rm atan}
 Q -1 \} $ \\
\\
$ \tilde j_2(ka \nu)={2 \over {1 +Q^2}}\{ 1 -{3 \over {2 Q^2}} ([Q +{1\over Q}]
{\rm atan} Q-1) \}         $ \\
\\
where $ Q={a\over 2} k \nu $ \\
\\
\hline
\end{tabular}
\end{center}
\vspace{2.0cm}
\begin{center}
{\bf Table IV}
 \end{center}
\begin{center}
\begin {tabular}{l|c|c|c|} \cline{2-4}
           &  $g$  & $h$ or $h'$  & size parameter (fm) \\ \hline
$\!\!\! \vline \ U(3)$     & 2.87 & -0.79 & $\alpha=$ 0.49 \\
$\!\!\! \vline \ SO(4)$     & 2.83  & -1.99 & $\beta= $ 0.54 \\
$\!\!\! \vline \ SO(4)^*$    & 2.96  & -1.91 & $a=$ 0.65  \\ \hline
\end{tabular}
\end{center}
\newpage
\begin{center}
{\bf Table V}
\end{center}
\begin{center}
\begin {tabular}{lrrrrr}
\hline \hline
Decay & Exp. value & $U(3)$ & $SO(4)$ & $SO(4)^*$ & $k\ \ $ \\
\hline
$\rho(770) \to \pi \pi $ & $151.5\pm 1.2 $ & $152.0$ & $151.8$ & $150.1$&
     $1.81$  \\
$\phi(1020) \to K^+  K^-$ & $2.18\pm 0.06$ &   $3.44$  & $3.39$ & $3.60$&
     $0.64$  \\
$ K^*(892) \to \pi K $   & $50.5\pm 0.6$ &   $48.5$  & $47.8$ & $49.8$ &
     $1.47$   \\
$ \phi(1020) \to \rho \pi $ & $0.57\pm 0.04$ & $0.46$  & $0.46$ & $0.48$ &
     $0.93$ \\
$ b_1(1235) \to \omega \pi $ & $155\pm 8$ & $82$ & $80$ & $83$ &
     $1.77$ \\
$K_1(1270) \to \rho K $ & $37.8\pm 13.8$ & $5.2$ & $5.2$ & $4.8$ &
     $0.36$  \\
$K_1(1270) \to K^* \pi $ & $14\pm 8$ & $51$ & $51$ & $51$ &
     $1.52$ \\
$a_1(1260) \to \rho \pi $ & $\sim 400\ \ \ $ & $341$ & $340$ & $341$ &
     $1.92$ \\
$K_1(1400) \to K^* \pi $ & $164\pm 23$ & $165$ & $170$ & $163$ &
     $2.03$ \\
$K_1(1400) \to \rho K $ & $5 \pm 5$ &
                     $63$ & $62$ & $62$ &  $1.52$ \\
$ a_2(1320) \to \eta \pi $ & $16\pm 2$ & $33$ & $30$ & $34$ &
     $2.71$  \\
$a_2(1320) \to K \bar K $ & $5.4\pm 1.1$ & $7.7$ & $6.9$ & $8.5$ &
     $2.21$  \\
$K^*_2(1430) \to \pi K $ & $54\pm 4$ & $36$ & $32$ & $38$ &
     $3.15$ \\
$K^*_2(1430) \to \eta  K $ & $0.15\pmatrix{+0.31 \cr -0.11}$ & $0.09$ &
$0.14$ & $0.05$&  $2.48$ \\
$f'_2(1525) \to K \bar K $ & $54 \pm 9$ & $43$ &
$41$ & $44$& $2.94$ \\
$f'_2(1525) \to \eta \eta  $ & $21 \pm 5$
                                     & $21$ & $20$ & $22$&
     $2.68$ \\
$f'_2(1525) \to \pi \pi $ & $0.62\pm 0.20$ & $8.94$ & $9.25$ & $8.06$&
     $3.80$ \\
$f_2(1270) \to \pi \pi $ & $157\pmatrix{+22 \cr -19}$& $151$ &
$146$ & $148$ &      $3.15$ \\
$f_2(1270) \to  K \bar K $ & $8.5\pm 1.9$ & $7.7$ & $6.9$ & $8.7$ &
     $2.04$ \\
$f_2(1270) \to\eta \eta  $ & $0.8\pm 0.3$ & $0.4$ & $0.3$ & $0.5$ &  $1.64$ \\
$K^*_0(1430) \to \pi K $ & $267\pm 50$ & $341$ & $339$ & $340$ &  $3.15$  \\
$a_2(1320) \to \rho \pi $ & $77\pm 6$ & $51$ & $45$ & $57$ &  $2.13$  \\
$K^*_2(1430) \to  K^* \pi $ & $27\pm 3$ & $28$ & $26$ & $29$ & $2.14$  \\
$K^*_2(1430) \to  \rho K $ & $9.6\pm 1.3$ & $3.5$ & $2.9$ & $4.2$ &   $1.69$ \\
$K^*_2(1430) \to  \omega  K $ & $3.16\pm 1.02$ & $1.15$ & $0.97$ & $1.39$&
     $1.62$  \\
\hline
\end{tabular}
\end{center}
\newpage
\begin{center} {\bf Table VI}
\end{center}
\begin{center}
\begin{tabular}{ll}
\hline
\hline
\\
$\Gamma_{a_0^+ \to \eta\, \pi^+}$ & $={k\over { 4 \pi}} \,  \chi \,
      [\sin\theta_P-{{\cos\theta_P}\over \sqrt{2}}]^2
           \vert F_5(1/2,k) \vert^2  $\\ \\
$\Gamma_{a_0^0 \to K^- K^+}$ & $=
    {{3k}\over{64\pi}}\,\vert F_5(1/2,k) \vert^2  $\\ \\
\hline
\end{tabular}
\end{center}
\begin{center}
\vspace{2.0cm}
{\bf Table VII}
\end{center}
\begin{center}
\begin{tabular}{l|c|c|c|}
        \cline{2-4}
      & \multicolumn{3}{|c|}{decay modes}   \\
        \cline{2-4}
      & $\rho \, \pi$ & $\eta \, \pi$ & $K\bar K$ \\ \hline
$\!\!\! \vline \ \,  a_2\, $ &  51        &  33        &  8	     \\
$\!\!\! \vline \ \, a_1\, $  &  341       &   0        &  0          \\
$\!\!\! \vline \ \, a_0\, $  &  0         &  335--700  &  86--240    \\
\hline
\end{tabular}
\end{center}

\vspace{2.0cm}

\newpage

\begin{thebibliography}{99}

\bibitem{PDB} Particle Data Book, Phys. Rev {\bf D45} (June 1992).

\bibitem{GM} M. Gell--Mann, Phys. Rev. {\bf 125}, 1067 (1962).

\bibitem{Nee} Y. Ne'eman, Nucl. Phys. {\bf 26}, 222 (1961).

\bibitem{QM} For a review, see J.J.J. Kokkedee, {\it The Quark Model}
(Benjamin, New York, 1969); and for recent calculations,
see N. Isgur and G. Karl, Phys. Rev. {\bf D18}, 4187 (1978),
and  Phys. Rev. {\bf D19}, 2653 (1979).

\bibitem{GI} S. Godfrey and N. Isgur, Phys. Rev. {\bf D32}, 189 (1985).

\bibitem{GR} F. G\"ursey and L.A. Radicati, Phys. Rev. Lett. {\bf 13},
173 (1964).

\bibitem{IM} F. Iachello, N.C. Mukhopadhyay and L. Zhang, Phys. Rev. {\bf D44},
898 (1991).

\bibitem{ID} F. Iachello and D. Kusnezov, Phys. Rev. {\bf D45}, 4156 (1992).

\bibitem{LeY} A. Le Yaouanc, LL. Oliver, O. P\`ene and J.-C~Raynal,
{\it Hadron Transitions in the Quark Model} (Gordon and Breach, New York,
1988).

\bibitem{JJS} J.J. de Swart, Rev. Mod. Phys. {\bf 35}, 916 (1963).

\bibitem{Glas} S.L. Glashow and A.H. Rosenfeld, Phys. Rev. Lett. {\bf 10},
192 (1963).

\bibitem{IL} F. Iachello and T.-S. Lee, preprint YCTP-N14-92 (1993).


\end{thebibliography}
\end{document}